\documentclass[10pt,technote]{article}
\usepackage[cp1250]{inputenc}
\usepackage{amsmath}
\usepackage{pst-plot}
\usepackage{pstricks}
\usepackage{pst-node}
\usepackage{amssymb}
\usepackage{listings}
\usepackage{url}

\title{Priority Queue based on multilevel prefix tree}
\author{David S. P\l aneta\footnote{dplaneta@gmail.com}}
\begin{document}
\maketitle
\begin{abstract}
Tree structures are very often used data structures. Among ordered types of 
trees there are many variants whose basic operations such as insert, delete, 
search, delete-min are characterized by logarithmic time complexity. 
In the article I am going to present the structure whose time complexity for 
each of the above operations is $O(\frac{M}{K} + K)$, where $M$ is the size 
of data type and $K$ is constant properly matching the size of data type. 
Properly matched $K$ will make the structure function as a very effective 
Priority Queue. The structure size linearly depends on the number and size 
of elements. PTrie is a clever combination of the idea of prefix tree -- 
Trie, structure of logarithmic time complexity for insert and remove 
operations, doubly linked list and queues.\break
\end{abstract}


\section{Introduction}
Priority Trie (PTrie~\cite{TR}) uses a few structures including Trie of 
$2^K$ degree~\cite{Bria}, which is the structure core. Data recording in 
PTrie consists in breaking the word into parts which make the indexes of the 
following layers in the structure (table look-at). The last layers contain 
the addresses of doubly linked list's nodes. Each of the list nodes stores 
the queue~\cite{Knuth1}, into which the elements are inserted. Moreover, each 
layer contains the structure of logarithmic time complexity of insert and 
remove operations. Which help to define the destination of data in the 
doubly linked list~\cite{Knuth1}. They can be various variants of ordered 
trees~\cite{Knuth1} or a skip list~\cite{Pugh}.
%
%
\subsection{Terminology}
Bit pattern is a set of $K$ bits. $K$ (length of bit pattern) defines the 
number of bits which are cut off the binary word. 
$M$ defines number (length) of bits in a binary word.
\begin{center}
value of word = $\overbrace{\underbrace{101...1}_{K}00101...}^{M}$
\end{center}
$N$ is  number of all values of PTrie. $2^K$ is variation $K$ of element 
binary set \{$0,1$\}. It determines the number of groups 
(number of Layers [Figure~\ref{Layer}]), which the bit pattern may be 
divided into during one step (one level). 
The set of values decomposed into the group by the first $K$
bits (the version of algorithm described in paper was implemented by 
machine of little-endian type).
\begin{figure}[!hbp]
\caption{Layer\label{Layer}}
\begin{center}
\begin{pspicture*}(0,4)(9,9.2)

\scriptsize
\rput[bc](6,8.4){\ovalnode{NodeA}{\textbf{A}}}
\rput[bc](7,7.4){\ovalnode{NodeC}{\textbf{C}}}
\rput[bc](5,7.4){\ovalnode{NodeB}{\textbf{B}}}
\rput[bc](4,6.4){\ovalnode{NodeD}{\textbf{D}}}
\ncline{<-}{NodeB}{NodeA}
\ncline{<-}{NodeC}{NodeA}
\ncline{<-}{NodeD}{NodeB}

\footnotesize
\rput[bl](6,6.5){ \textbf{\normalsize{\ldots}} }

\psline{|-|}(7.6,9)(7.6,6)
\rput[bl](7.8,6){\rotateright{$log_{2}P = log_{2}2^K = K$}}

\pscurve(3.6,6)(3.5,6)(3.2,6.4)(3.2,7)(3,7.5)%
        (3,7.5)(3.2,8)(3.2,8.6)(3.5,9)(3.6,9)
\rput[bl](0.2,6.8){\parbox{2.7cm}{The Structure of logarithmics time complexity of insert and remove operations}}

 \rput[bl](0.2,5){\rnode{BlockMIN}{\psdblframebox{ \parbox{18pt}{MIN} }}}
 \rput[bl](1.6,5){\rnode{BlockMAX}{\psdblframebox{ \parbox{18pt}{MAX} }}}
 
 \rput[bl](3,5){\rnode{blockA}{\psframebox{ $00....00 $ }}}
 \rput[bl](3.6,5.6){$G_{1}$}
 
 \rput[bl](4.8,5){\rnode{blockB}{\psframebox{ $ 00...01 $ }}}
 \rput[bl](5.4,5.6){$G_{2}$}
 
 \rput[bl](6.2,5){ \textbf{\normalsize{\ldots}} }
 
 \rput[bl](7,5){\rnode{blockC}{\psframebox{ $ 11...11 $ }}}
 \rput[bl](7.6,5.6){$G_{P}$}
 
 \psline{|-|}(2.9,4.6)(8.4,4.6)
 \rput[bc](5.6, 4.4) {$P=2^K$}
 
\end{pspicture*}
\end{center}
\end{figure}
The path is defined starting from the most important bits of variable. 
The value of pattern $K$ (index) determines the layer we move to 
[Figure~\ref{PTrie}]. The lowest layers determine the nodes of the 
list which store the queues for inserted values. L defines the level 
the layer is on. 
\begin{figure}[!hbp]
\begin{center}
\caption{PTrie\label{PTrie}}
\begin{pspicture*}(-1,-1)(11,9.4)
\scriptsize
 \rput[bc](6,8.4){\ovalnode{NodeA}{\textbf{A}}}
 \rput[bc](7,7.4){\ovalnode{NodeC}{\textbf{C}}}
 \rput[bc](5,7.4){\ovalnode{NodeB}{\textbf{B}}}
 \rput[bc](4,6.4){\ovalnode{NodeD}{\textbf{D}}}
 \ncline{<-}{NodeB}{NodeA}
 \ncline{<-}{NodeC}{NodeA}
 \ncline{<-}{NodeD}{NodeB}

\footnotesize
 \rput[bl](6,6.5){ \textbf{\normalsize{\ldots}} }

 \psline{|-|}(7.6,9)(7.6,6)
 \rput[bl](7.8,6.4){\rotateright{$\Theta(log_{2}2^K) = O(K)$}}

 \rput[bl](0.2,8.8){\textbf{Layer}}

 \rput[bl](0.2,5){\rnode{BlockMIN}{\psdblframebox{ \parbox{18pt}{MIN} }}}
 \rput[bl](1.6,5){\rnode{BlockMAX}{\psdblframebox{ \parbox{18pt}{MAX} }}}
 
 \rput[bl](3,5){\rnode{BlockA}{\psframebox{ $00....00 $ }}}
 \rput[bl](3.6,5.6){$G_{1}$}
 
 \rput[bl](4.8,5){\rnode{BlockB}{\psframebox{ $ 00...01 $ }}}
 \rput[bl](5.4,5.6){$G_{2}$}
 
 \rput[bl](6.2,5){ \rnode{Block...}{\textbf{\normalsize{\ldots}} }}
 
 \rput[bl](7,5){\rnode{BlockC}{\psframebox{ $ 11...11 $ }}}
 \rput[bl](7.6,5.6){$G_{P}$}
 
 \psline{|-|}(2.9,4.6)(8.4,4.6)
 \rput[bc](5.6, 4.4) {$P=2^K$}
 
 \psframe[dimen=inner](0,4)(9,9.2)

\psframe[dimen=inner](0,2)(2,3)
\rput(1,2.5){\rnode{LayerA}{\textbf{Layer}}}
\ncline{<-}{LayerA}{BlockA}

\psframe[dimen=inner](3,2)(5,3)
\rput(4,2.5){\rnode{LayerB}{\textbf{Layer}}}
\ncline{<-}{LayerB}{BlockB}

\psframe[dimen=inner](8,2)(10,3)
\rput(9,2.5){\rnode{LayerC}{\textbf{Layer}}}
\ncline{<-}{LayerC}{BlockC}

\rput[bl](6.2,2.2){ \rnode{Layer...}{\textbf{\normalsize{\ldots}} }}
\ncline{<-}{Layer...}{Block...}

\rput[bl](0.5,1.5){ \rnode{dots1}{\textbf{\normalsize{\ldots}} }}
\rput[bl](2,1.5){ \rnode{dots2}{\textbf{\normalsize{\ldots}} }}
\rput[bl](3,0.2){ \rnode{dots3}{\textbf{\normalsize{\ldots}} }}
\rput[bl](3,1.5){ \rnode{dots4}{\textbf{\normalsize{\ldots}} }}
\rput[bl](4.1,1.5){ \rnode{dots5}{\textbf{\normalsize{\ldots}} }}
\rput[bl](6.2,0.2){ \rnode{dots6}{\textbf{\normalsize{\ldots}} }}
\rput[bl](8.6,1.5){ \rnode{dots7}{\textbf{\normalsize{\ldots}} }}

\ncline{<-}{dots1}{LayerA}
\ncline{<-}{dots2}{LayerA}
\ncline{<-}{dots3}{dots2}
\ncline{<-}{dots4}{LayerB}
\ncline{<-}{dots3}{dots4}
\ncline{<-}{dots5}{LayerB}
\ncline{<-}{dots6}{Layer...}
\ncline{<-}{dots7}{LayerC}

\psline{|-|}(10.2,9.4)(10.2,1)
\rput[bl](10.3,3){\rotateright{$\Theta(log_{2^K}N) = \Theta(\frac{lgN}{lg2^K}) = O(\frac{M}{K})$}}
 
\rput[bl](-0.8,6.5){$L_{1}$}
\rput[bl](-0.8,2.5){$L_{2}$}
\rput[bl](-0.8,1.4){$L_{log_{2^K}N}$}
\rput[bl](0,0){\rnode{Node1}{\psframebox{$Node$}}}
\ncline{<-}{Node1}{dots1}
\rput[bl](0,-1){\rnode{Q1}{\psframebox{\scriptsize{Queue}}}}
\ncline{<-}{Q1}{Node1}

\rput[bl](1.5,0){\rnode{Node2}{\psframebox{$Node$}}}
\ncline{<-}{Node2}{dots1}
\rput[bl](1.5,-1){\rnode{Q2}{\psframebox{\scriptsize{Queue}}}}
\ncline{<-}{Q2}{Node2}

\rput[bl](4.2,0){\rnode{Node3}{\psframebox{$Node$}}}
\ncline{<-}{Node3}{dots5}
\rput[bl](4.2,-1){\rnode{Q3}{\psframebox{\scriptsize{Queue}}}}
\ncline{<-}{Q3}{Node3}

\rput[bl](7.5,0){\rnode{Node4}{\psframebox{$Node$}}}
\ncline{<-}{Node4}{dots7}
\rput[bl](7.5,-1){\rnode{Q4}{\psframebox{\scriptsize{Queue}}}}
\ncline{<-}{Q4}{Node4}

\rput[bl](9,0){\rnode{Node5}{\psframebox{$Node$}}}
\ncline{<-}{Node5}{dots7}
\rput[bl](9,-1){\rnode{Q5}{\psframebox{\scriptsize{Queue}}}}
\ncline{<-}{Q5}{Node5}

\rput[tl](10.5,1){\rnode{Tail}{\rotateright{\psframebox{Tail}} }}
\nccurve[angleB=270]{<-}{Node5}{Tail}

\rput[tl](-1,1){\rnode{Head}{\rotateright{\psframebox{Head}} }}
\nccurve[angleB=180, angleA=270]{->}{Head}{Node1}

\ncarc{->}{Node1}{Node2}
\ncarc{->}{Node2}{Node1}

\ncarc{->}{dots3}{Node2}
\ncarc{->}{Node2}{dots3}

\ncarc{->}{dots3}{Node3}
\ncarc{->}{Node3}{dots3}

\ncarc{->}{dots6}{Node3}
\ncarc{->}{Node3}{dots6}

\ncarc{->}{dots6}{Node4}
\ncarc{->}{Node4}{dots6}

\ncarc{->}{Node5}{Node4}
\ncarc{->}{Node4}{Node5}

\end{pspicture*}
\end{center}
\end{figure}
Probability that exactly $G$ keys correspond to one particular pattern, 
where for each of $P_{L}$ sequences of leading bits there is such a node 
that corresponds to at least two keys equals
\begin{center}
 ${N \choose G} P^{-GL}(1-P^{-L})^{N-G}$
\end{center}
For random PTrie the average number of layers on level $L$, for 
$L = 0, 1, 2, \ldots $is
\begin{center}
 $P^L(1-(1-P^{-L})^N)-N(1-P^{-L})^{N-1}$
\end{center}
If $A_{N}$ is average number of layers in random PTrie of degree $P=2^K$ 
containing $N$ keys. Then $A_{0} = A_{1} = 0$, and for $N \geq 2$ 
we get~\cite{Knuth}:
\begin{center}
 \begin{displaymath}
  A_{N} = 1 + \sum_{G_{1} + \ldots + G_{P} = N} 
  \Big(\frac{N!}{G_{1}! \ldots G_{P}!} P^{-N}\Big) 
  \Big(A_{G_{1}} + \ldots + A_{G_{P}}\Big) = 
 \end{displaymath}
 \begin{displaymath}
  1 + P^{1-N} \sum_{G_{1} + \ldots + G_{P} = N}
  \Big(\frac{N!}{G_{1}! \ldots G_{P}!}\Big) A_{G_{1}} =
 \end{displaymath}
 \begin{displaymath}
  1 + P^{1-N} \sum_{G}{N \choose G} \Big(P - 1 \Big)^{N-G} A_{G} =
 \end{displaymath}
  \begin{displaymath}
  1 + 2^{G(1-N)} \sum_{G} {N \choose G} \Big( 2^G - 1 \Big)^{N-G} A_{G}
 \end{displaymath}
\end{center}
\section{Implementation}
\begin{center}
\begin{tabular}{|l|l|l|}
\hline
\textbf{Operation} & \textbf{Description} & \textbf{Bound}\\
\hline
create & Creates object & $O(1)$ \\
\hline
insert(data) & Adds element to the structure. & $O(\frac{M}{K} + K)$\\
\hline
boolean remove(data) & 
\parbox{5.5cm}{Removes value from the tree. If operation failed because there 
was no such value in the tree it returns FALSE(0), otherwise returns TRUE(0).}
& $O(\frac{M}{K} + K)$\\
\hline
boolean search(data) &
\parbox{5.5cm}{
Looks for the words in the tree. If finds return TRUE(1), 
otherwise FALSE(0).}
& $O(\frac{M}{K})$\\
\hline
*minimum() & 
\parbox{5.5cm}{
Returns the address of the lowest value in the tree, or empty address if the 
operation failed because the tree was empty.
}
& $O(1)$\\
\hline
*maximum() & 
\parbox{5.5cm}{
Returns the address of the highest value in the tree or empty address if the 
operation failed because the tree was empty.
}
& $O(1)$\\
\hline
next & 
\parbox{5.5cm}{
Returns the address of the next node in the tree or empty address if value 
transmitted in parameter was the greatest. The order of moving to successive 
elements is fixed - from the smallest to the largest and from 
``the youngest to the oldest'' (stable) in case of identical words.
}
& $O(1)$\\
\hline
back &
\parbox{5.5cm}{
Similar to `next' but it returns the address of preceding node in the tree.
}
& $O(1)$\\
\hline
\end{tabular}
\end{center}
Basic operations can be joined. For example, the effect connected with the 
heap; delete-min() can be replaced by operations remove(minimum()).
\subsection{Insert}
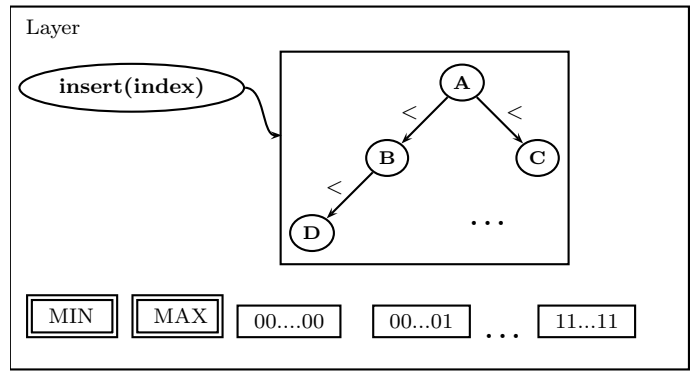
\begin{figure}[!hbp]
\caption{Insert value of index into the ordered binary tree from the layer\label{Tree}}
\begin{center}
\begin{pspicture*}(0,4.5)(9.2,9.5)

\scriptsize
\rput[bc](6,8.4){\ovalnode{NodeA}{\textbf{A}}}
\rput[bc](7,7.4){\ovalnode{NodeC}{\textbf{C}}}
\rput[bc](5,7.4){\ovalnode{NodeB}{\textbf{B}}}
\rput[bc](4,6.4){\ovalnode{NodeD}{\textbf{D}}}
\ncline{<-}{NodeB}{NodeA}
\ncline{<-}{NodeC}{NodeA}
\ncline{<-}{NodeD}{NodeB}

\footnotesize
\psframe[dimen=inner](3.6,6)(7.4,8.8)
\rput[bl](3.6,7.7){\rnode{square}{}}
\rput[bl](0.2,9){Layer}
\rput[bl](0.1,8){\ovalnode{insert}{\textbf{insert(index)}}}
\nccurve[angleB=180]{->}{insert}{square}

\rput[bc](5.3,8){ \textbf{$<$} }
\rput[bc](6.7,8){ \textbf{$<$} }
\rput[bc](4.3,7){ \textbf{$<$} }

\rput[bl](6,6.5){ \textbf{\normalsize{\ldots}} }

 \rput[bl](0.2,5){\rnode{BlockMIN}{\psdblframebox{ \parbox{18pt}{MIN} }}}
 \rput[bl](1.6,5){\rnode{BlockMAX}{\psdblframebox{ \parbox{18pt}{MAX} }}}
 \rput[bl](3,5){\rnode{blockA}{\psframebox{ $00....00 $ }}}
 \rput[bl](4.8,5){\rnode{blockB}{\psframebox{ $ 00...01 $ }}}
 \rput[bl](6.2,5){ \textbf{\normalsize{\ldots}} }
 \rput[bl](7,5){\rnode{blockC}{\psframebox{ $ 11...11 $ }}}

\psframe[dimen=inner](0,4.6)(9,9.4)
\end{pspicture*}
\end{center}
\end{figure}

Determine the interlinked index (pointer) to another layer using the length 
of pattern projecting on the word.\\
\indent\emph{\textbf{If}} interlink determined by index is not empty and indicated 
the list node -- try to insert the value into the queue of determined node.\\
\emph{If} the elements in the queue turn out to be the same,  
insert value into the queue.
\emph{Otherwise}, if elements in the queue are different from 
the inserted value, the node is ``pushed'' to a lower level and the hitherto 
existing level (the place of node) is complemented with a new layer. 
Next, try again to insert the element, this time however, into the newly 
created layer.\\
\indent\emph{\textbf{Else}}, if the interlink determined by index is empty, insert 
value of index into the ordered binary tree from the current layer 
[Figure~\ref{Tree}]. Father of a newly created node in ordered binary 
tree from the current layer determines the place for leaves; If the newly 
created node in ordered binary tree is on the right side of father 
(added index $>$ father index), the value added to the list will be 
inserted after the node determined by father index and the path of the 
highest indexes (make use of pointer `max' of the layers -- time cost $O(1)$) 
of lower level layers. If newly created node is on the left side of father 
(added index $<$ father index), the value added to the list will be 
inserted before the node determined by father index and the path of the 
smallest indexes (make use of pointer `min' of the layers -- time cost $O(1)$) 
of lower level layers.\break
\noindent%
One can wonder why we use the queue and not the stack or the value counter. 
Value counter cannot be used because complex elements can be inserted into 
PTrie structure, distinguishable in the tree only because of some words. 
Also, it is not a good idea to use a stack because the queue makes the 
structure stable. And this is a very useful characteristic.
I used ``plain'' Binary Search Tree in the structure of logarithmic time 
complexity. For a small number of tree nodes it is a very good solution 
because for $K = 4$, $2^K = 16$. So in the tree there may be maximum $16$ 
(different) elements. For such a small amount of (different) values the 
remaining ordered trees will probably turn out to be at most as effective 
as unusually simple Binary Search Trees.\break
%
%
\subsubsection{Analysis}
In case of random data it will take 
$\Theta(\frac{lgN}{lg2^K}) = \Theta(log_{2^K}N) = O(\frac{M}{K})$ goings 
through layers to find the place in the heap core -- Trie tree. 
On at least one layer of PTrie structure we will use inserting into the 
ordered binary tree in which maximum number of nodes is $2^K$. 
While inserting the new value I need information where exactly it will be 
located in the list. Such information can be obtained in two ways; 
I will get the information if the representation of the nearest index on the 
list is to the left or to the right side of the inserted word index. 
It may happen that in the structure there is already is exactly the same 
word as the inserted one. In such case value index won't be inserted into 
any layer of the PTrie because it will not be necessary to add a new 
node of the list. Value will be inserted into the queue of already 
existing node. To sum up, while moving through the layers of PTrie we can 
stop at some level because of empty index. Then, a node will be added to 
the list in place determined by binary search tree and the remaining 
part of the path. This is why the bound of operation which inserts new 
value into PTrie equals
$\Theta(log_{2^K}N +log_{2}{2^K}) = \Theta(log_{2^K}N + K) = O(\frac{M}{K} + K)$.
\break
%
%
\subsection{Find}
Method find like in case of plain Trie trees goes through succeeding layers 
following the path determined by binary representation of search value. 
It can be stated that it uses number key as a guide while moving down the 
core of PTrie -- prefix tree. In case of searching tree things can happen:
\begin{itemize}
	\item We don't reach the node of the list because the index we determine 
	      is empty on any of layers -- searching failure.
	\item We reach the node but values from the queue are different from the 
	      searched value -- searching failure.
	\item We reach the node and the values from the queue are exactly like 
	      the ones we seek -- searching success.
\end{itemize}
%
%
\subsubsection{Analysis}
Searching in prefix tree is very fast because it finds the words using word 
key as indexes. In case of search failure the longest match of a searched 
word is found. It must be taken into consideration that during operation 
`search' we use only the attributes of prefix tree. This is why the amount 
of search numbers looked through during the random search is 
$\Theta(log_{2^K}N) = O(\frac{M}{K})$.\break
%
%
\subsection{Remove}
Remove method just like find method ``moves down'' the PTrie structure to 
seek for the element to be deleted. If it doesn't reach the node of 
the list, or it does but the search value is different from the value of 
node queue, it does not delete any element of PTrie because it is not there. 
However if it reaches the node of the list and search value turns out to be 
the value from the queue -- it removes the value from the queue. 
If it remains empty after removing the element from the queue the node 
will be removed from the list and will return to the ``upper'' layers 
of prefix tree to delete possible, remaining, empty layers.\break
%
%
\subsubsection{Analysis}
Since it is possible not only to go down the tree but also come back upwards 
(in case of deleting of the lower layer or the node of the list) the total 
length of the path move on is limited $\Theta(2log_{2^K}N)$. If delete 
the layer, it means there was only one way down from that layer, which 
implicates the fact that the ordered binary tree of a given layer contained 
only one node (index). The layer is removed if it remains empty after 
the removal of node from ordered binary tree. So the number of operation 
necessary for the removal of the layer containing one element equals 
$\Theta(1)$. In case of removal of layer $L_{i}$, if ordered binary tree 
of higher level layer $L_{i-1}$, despite removing the node which determines 
empty layer we came from, does not remain empty it means that there could be 
maximum $2^K$ nodes in the ordered binary tree. Operation of value delete 
from ordered binary tree amounts to $\Theta(log_{2}2^K) = \Theta(K)$. 
There is no point of ``climbing'' up the upper layers, since the layer we 
came from would not be empty. At this stage the method remove ends. 
To sum up, worse time complexity of remove operation is 
$\Theta(2log_{2^K}N + K) = O(\frac{M}{K} + K)$.\break
%
%
\subsection{Extract minimum and maximum}
If the list is not empty, `minimum' reads the value pointed by the head of 
the list and `Maximum' reads the value pointed by the tail of the list.
%
%
\subsubsection{Analysis}
Time complexity of operations is $\Theta(1)$.
%
%
\subsection{Iterators}
The nodes of the list are linked. If we know the position of one of the nodes, 
we have a direct access to its neighbors. The `next' operation reads the 
successor of current pointed node. The `prev' operation reads the predecessor 
of currently pointed node.
%
%
\subsubsection{Analysis}
Moving to the node its neighbor requires only reading of the contents of the 
pointer `next' or `prev'. Time complexity of such operations equals 
$\Theta(1)$.
\section{Correctness}
PTrie has been designed like this, so as not to assume that keys have to be 
positive numbers or only integers - they can be even strings (however, 
in most cases the weight of arcs is represented by numbers).
To insert PTrie negative and positive integers I use not one PTrie, but two! 
One of the structures is destined exclusively for storing positive integers 
and the other one for storing only negative integers. The latter structure 
of PTrie is responsible only for negative integers - the integers are stored 
in reverse order on the list (for machine of little-endian type).
Therefore in case of the second structure of PTrie (responsible only for 
negative integers) I used standard operation of PTrie: PTrie2.maximum to 
extract the smallest value. Also real numbers (for example in 
ANSI IEEE 754-1985 standard) can be used of the 
description of the weight of arcs on condition that two 
interrelated structures of PTrie will be used to put off exponent and 
mantissa. It is possible, because implementation of PTrie
described by me uses queue, which makes it stable. One of the structures 
of PTrie serves as storage for exponent, where each of the nodes of the 
list will contain additional structure of PTrie to store mantissa.
\section{Conclusions}
Efficiency of PTrie (source codes~\cite{TR}) considerably depends on the length of pattern $K$. 
$K$ defines optional value, which is the power of two in the range 
[$1$, min($M$)]. The total size of necessary memory bound is proportional to 
$\Theta(\frac{log_{2^K}N(2^{K+1})}{K})$ because the number of layers 
required to remember $N$ random elements in PTrie of degree $2^K$ equals 
$\frac{lgN}{lgP}*P$. Moreover, each layers has tree of maximum size $2^K$ 
nodes and table of the $P$-elements, so the necessary memory bound equal 
$\Theta(log_{2^K}N * 2P) = \Theta(\frac{M}{K} * 2^{K+1})$. 
For data types of constant size maximum Trie tree height equals 
$\frac{M}{K}$. So the pessimistic operation time complexity is 
$O(\frac{M}{K} + K)$. For example, for four-byte numbers it is the most 
effective to determine the pattern $K=4$ bits long. Then, the pessimistic 
number of steps necessary for the operation on the PTrie will equal 
$\Theta(\frac{M}{K} + K) = \frac{32}{4} + 4 = 12$. 
Increasing $K$ to $K=8$ does not increase the efficiency of the structure 
operation because $\Theta(\frac{M}{K} + K) = \frac{32}{8} + 8 = 12$. 
What is more, in will unnecessarily increase the memory demand. A single 
layer consisting of $P=2^K$ groups for $K=8$ will contain tables 
$P=2^8=256$ long, not when $K=4$, only $P=2^4=16$ links. For variable size 
data the time complexity equals $\Theta(log_{2^K}N + K)$. Moreover, the 
length of pattern $K$ must be carefully matched. For example, for strings $K$ 
should not be longer than $8$ bits because we could accidentally read the 
contents from beyond the string which normally consist of one-byte sign! 
It is possible to record data of variable size in the structure provided 
each of the analyzed words will end with identical key. There are no 
obstacles for strings because they normally finish with ``end of line'' sign.
Owing to the reading of word keys and going through indexes (table look-at), 
primary, partial operations of PTrie method are very fast. If we carefully 
match $K$ with data type, PTrie will certainly serve as a really effective 
Priority Queue~\cite{GRAPH}.\break

%
%
%
\end{document}